# Cloudbus Toolkit for Market-Oriented Cloud Computing

Rajkumar Buyya[1,2], Suraj Pandey[1], and Christian Vecchiola[1]

[1]Cloud Computing and Distributed Systems (CLOUDS) Laboratory
Department of Computer Science and Software Engineering
The University of Melbourne, Australia
Email: {raj, spandey, csve}@csse.unimelb.edu.au

[2]Manjrasoft Pty Ltd, Melbourne, Australia

**Abstract.** This keynote paper: (1) presents the 21st century vision of computing and identifies various IT paradigms promising to deliver computing as a utility; (2) defines the architecture for creating market-oriented Clouds and computing atmosphere by leveraging technologies such as virtual machines; (3) provides thoughts on market-based resource management strategies that encompass both customer-driven service management and computational risk management to sustain SLA-oriented resource allocation; (4) presents the work carried out as part of our new Cloud Computing initiative, called Cloudbus: (i) Aneka, a Platform as a Service software system containing SDK (Software Development Kit) for construction of Cloud applications and deployment on private or public Clouds, in addition to supporting market-oriented resource management; (ii) internetworking of Clouds for dynamic creation of federated computing environments for scaling of elastic applications; (iii) creation of $3^{rd}$ party Cloud brokering services for building content delivery networks and e-Science applications and their deployment on capabilities of IaaS providers such as Amazon along with Grid mashups; (iv) CloudSim supporting modelling and simulation of Clouds for performance studies; (v) Energy Efficient Resource Allocation Mechanisms and Techniques for creation and management of Green Clouds; and (vi) pathways for future research.

## 1 Introduction - Technology Trends

In 1969, Leonard Kleinrock, one of the chief scientists of the original Advanced Research Projects Agency Network (ARPANET) project which seeded the Internet, said [1]: *"As of now, computer networks are still in their infancy, but as they grow up and become sophisticated, we will probably see the spread of 'computer utilities' which, like present electric and telephone utilities, will service individual homes and offices across the country."* This vision of computing utilities, based on a service provisioning model, anticipated the massive transformation of the entire computing industry in the $21^{st}$ century whereby computing services will be readily available on demand, like water, electricity, gas, and telephony services available in today's society. Similarly, computing service users (consumers) need to pay providers only when they access computing services, without the need to invest heavily or encounter difficulties in building and maintaining complex IT infrastructure by themselves. They access the services based on their requirements without regard to where the services are hosted. This model has been referred to as *utility computing*, or recently as *Cloud computing* [2].

Cloud computing delivers infrastructure, platform, and software (application) as services, which are made available as subscription-based services in a pay-as-you-go model to consumers. These services in industry are referred to as Infrastructure as a Service (IaaS), Platform as a Service (PaaS), and Software as a Service (SaaS), respectively. Berkeley Report [3] released in Feb 2009 notes - "Cloud computing, the long-held dream of computing as a utility, has the potential to transform a large part of the IT industry, making software even more attractive as a service".

Clouds aim to power the next generation data centers by architecting them as a network of virtual services (hardware, database, user-interface, application logic) so that users are able to access and deploy applications from anywhere in the world on demand at competitive costs depending on users Quality of Service (QoS) requirements [4]. It offers significant benefit to IT companies by freeing them from the low level tasks of setting up basic hardware (servers) and software infrastructures and thus enabling them to focus on innovation and creating business value for their services.

The business potential of Cloud computing is recognised by several market research firms including IDC (International Data Corporation), which reports that worldwide spending on Cloud services will grow from $16 billion by 2008 to $42 billion in 2012. Furthermore, many applications making use of Clouds emerge simply as catalysts or market makers that bring buyers and sellers together. This creates several trillion dollars of business opportunity to the utility/pervasive computing industry, as noted by Bill Joy, co-founder of Sun Microsystems [5].

Cloud computing has high potential to provide infrastructure, services and capabilities required for harnessing this business potential. In fact, it has been identified as one of the emerging technologies in IT as noted in "Gartner's IT Hype Cycle" (see Figure 1). A "Hype Cycle" is a way to represent the emergence, adoption, maturity and impact on applications of specific technologies.

Cloud computing is definitely at the top of the technology trend, reaching its peak of expectations in just 3-5 years. This trend is enforced by providers such as Amazon[1], Google, SalesForce[2], IBM, Microsoft, and Sun Microsystems who have begun to establish new data centers for hosting Cloud computing applications such as social networking (e.g. Facebook[3] and MySpace[4]), gaming portals (e.g. BigPoint[5]), business applications (e.g., SalesForce.com), media content delivery, and scientific workflows. It is predicted that within the next 2-5 years, Cloud computing will become a part of mainstream computing; that is, it enters into the plateau of productivity phase.

---

[1] http://www.amazon.com/
[2] http://www.salesforce.com/
[3] http://www.facebook.com/
[4] http://www.myspace.com/
[5] http://www.bigpoint.net/

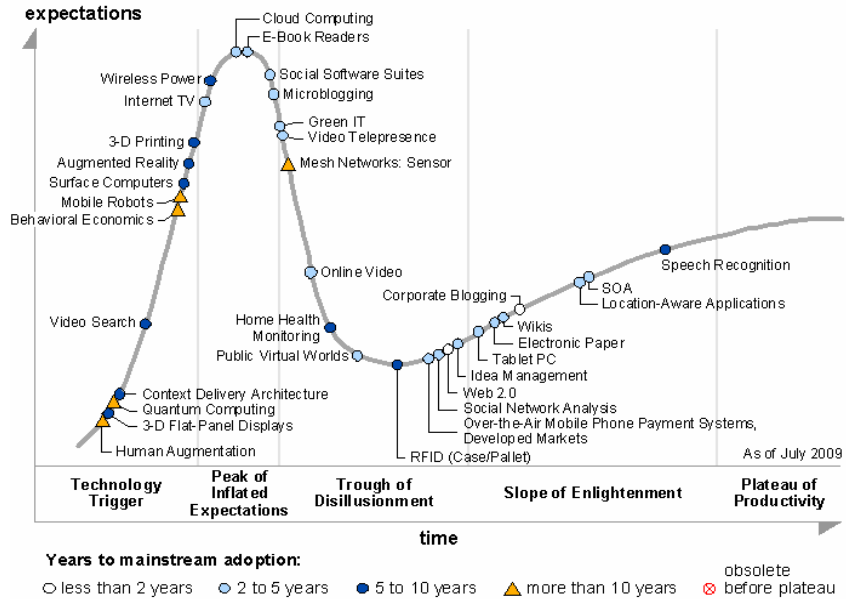

**Fig. 1.** Hype Cycle of Emerging Technologies, 2009 - Source: Gartner (August 2009).

The rest of the paper is organized as follows: Section 2 presents a high-level definition of Cloud computing followed by open challenges and a reference model; Section 3 presents Cloudbus vision and architecture in conformance with the high-level definition; Section 4 lists specific technologies of the Cloudbus toolkit that have made the vision a reality; Section 5 talks about integration of the Cloudbus toolkit with other Cloud management Technologies; and finally, Section 6 concludes the paper providing insights into future trends in Cloud computing.

## 2 Cloud Computing

Cloud computing [3] is an emerging paradigm that aims at delivering hardware infrastructure and software applications as services, which users can consume on a pay-per-use-basis. As depicted in Fig. 1, Cloud computing is now at the peak of its hype cycle and there are a lot of expectations from this technology. In order to fully understand its potential, we first provide a more precise definition of the term, then introduce a reference model for Cloud computing, and briefly sketch the challenges that lies ahead.

## 2.1 Cloud Definition

The cloud symbol traditionally represents the Internet. Hence, Cloud computing refers to the practice of moving computing to the Internet. Armbrust et al. [3] observe that *"Cloud computing refers to both the applications delivered as services over the Internet and the hardware and system software in the data centers that provide those services"*. This definition captures the real essence of this new trend, where both software applications and hardware infrastructures are moved from private environment to third parties data centers and made accessible through the Internet. Buyya et al. [2] define Cloud as *"a type of parallel and distributed system consisting of a collection of interconnected and virtualized computers that are dynamically provisioned and presented as one or more unified computing resources based on service-level agreements"*. This definition puts Cloud computing into a market oriented perspective and stresses the economic nature of this phenomenon.

The key feature, emerging from above two characterizations is the ability to deliver both infrastructure and software as services that are consumed on a pay-per-use-basis. Previous trends were limited to a specific class of users, or specific kinds of IT resources; the approach of Cloud computing is global and encompasses the entire computing stack. It provides services to the mass, ranging from the end-users hosting their personal documents on the Internet to enterprises outsourcing their entire IT infrastructure to external data centers. Service Level Agreements (SLAs), which include QoS requirements, are set up between customers and Cloud providers. An SLA specifies the details of the service to be provided in terms of metrics agreed upon by all parties, and penalties for violating the expectations. SLAs act as a warranty for users, who can more comfortably move their business to the Cloud. As a result, enterprises can cut down maintenance and administrative costs by renting their IT infrastructure from Cloud vendors. Similarly, end-users leverage the Cloud not only for accessing their personal data from everywhere, but also for carrying out activities without buying expensive software and hardware.

Figure 2 shows the high level components of the service-oriented architectural framework consisting of client's brokering and coordinator services that support utility-driven management of Clouds: application scheduling, resource allocation and migration of workloads. The architecture cohesively couples the administratively and topologically distributed storage and compute capabilities of Clouds as parts of a single resource leasing abstraction [4]. The system will ease the cross-domain integration of capabilities for on-demand, flexible, energy-efficient, and reliable access to the infrastructure based on emerging virtualization technologies [6,7].

The Cloud Exchange (CEx) acts as a market maker for bringing together service producers and consumers. It aggregates the infrastructure demands from the application brokers and evaluates them against the available supply currently published by the Cloud Coordinators. It aims to support trading of Cloud services based on competitive economic models such as commodity markets and auctions. CEx allows the participants (Cloud Coordinators and Cloud Brokers) to locate providers and consumers with fitting offers. Such markets enable services to be commoditized and thus, can pave the way for the creation of dynamic market infrastructure for trading based on SLAs. The availability of a banking system within the market ensures that financial transactions pertaining to SLAs between participants are carried out in a secure and dependable environment. Every client in the Cloud platform will need to instantiate a Cloud brokering service that can dynamically establish service contracts with Cloud Coordinators via the trading functions exposed by the Cloud Exchange.

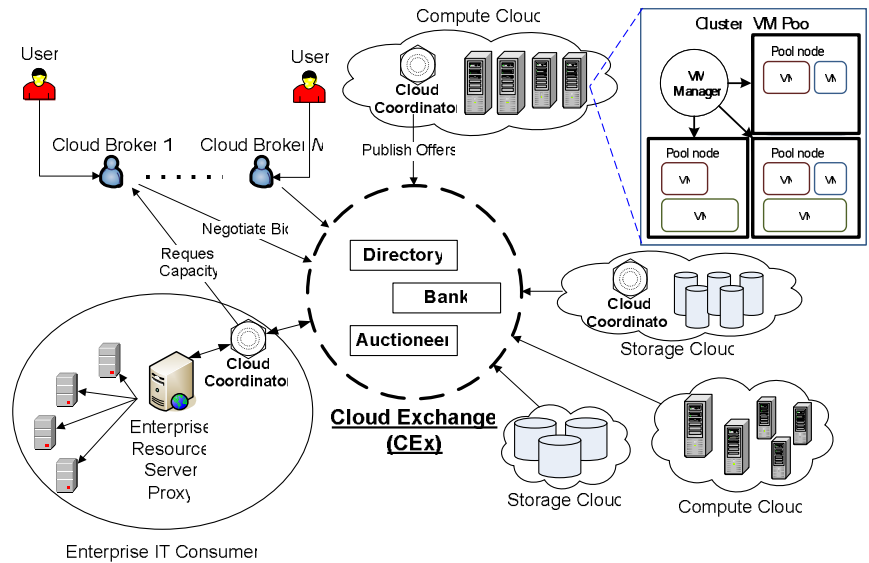

**Fig. 2.** Utility-oriented Clouds and their federated network mediated by Cloud exchange.

### 2.2 Open Challenges

Cloud computing introduces many challenges for system and application developers, engineers, system administrators, and service providers. Fig. 3 identifies some of them. Virtualization enables consolidation of servers for hosting one or more services on independent virtual machines in a multi-tenancy manner. When a large number of VMs are created they need to be effectively managed to ensure that services are able to deliver quality expectations of users. That means, VMs need to be migrated to suitable servers when QoS demand on services is high and later get consolidated dynamically to a fewer number of physical servers.

One of the major concerns when moving to Clouds is related to security, privacy, and trust. Security in particular, affects the entire cloud computing stack. The Cloud computing model promotes massive use of third party services and infrastructures to host important data or to perform critical operations. In this scenario, the trust towards providers is fundamental to ensure the desired level of privacy for applications hosted in the Cloud. At present, traditional tools and models used to enforce a secure and reliable environment from a security point of view are the only ones available.

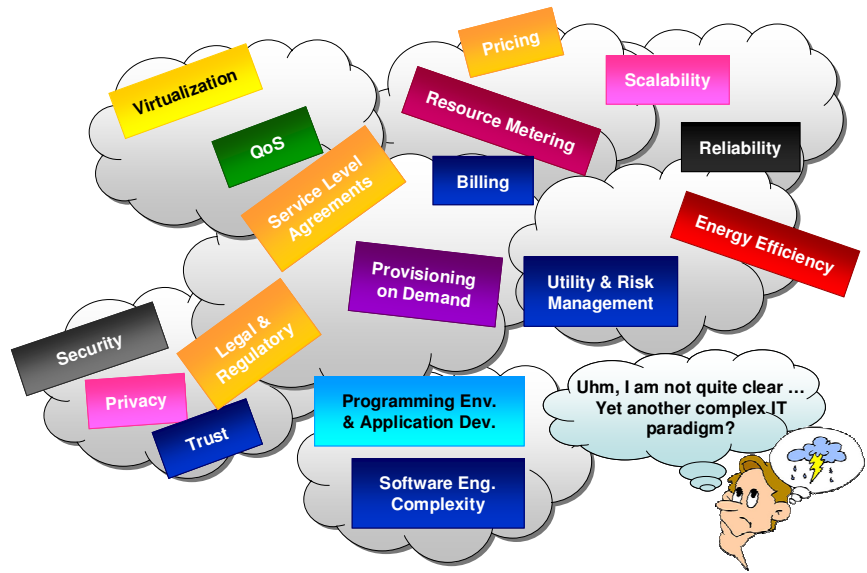

**Fig. 3.** Cloud computing challenges.

Besides security, there are legal and regulatory issues that need to be taken care of. When moving applications and data to the Cloud, the providers may choose to locate them anywhere on the planet. The physical location of data centers and clusters determines the set of laws that can be applied to the management of data. For example, specific cryptography techniques could not be used because they are not allowed in some countries. Simply, specific classes of users, such as banks, would not be comfortable to put their sensitive data into the Cloud, in order to protect their customers and their business. At present, a conservative approach is taken for what concerns hosting sensitive data. An interesting initiative is the concept of availability zones[6] promoted by Amazon EC2. Availability zones identify a set of resources that have a specific geographic location. Currently there are two regions grouping the availability zones: US and Europe. Although this initiative is mostly concerned with providing of better services in terms of isolation from failures, network latency, and service downtime, it could be an interesting example for exploring legal and regulatory issues.

In most cases, the desired level of security is established within a Service Level Agreement (SLA). SLAs also establish the price of services, and specific activities such as resource metering, billing, and pricing have to be implemented in order to charge users. At present, the adopted solutions fall into the "pay-as-you-go" model, where users are charged according to the use they make of the service. More sophisticated and flexible pricing policies have to be developed and put in place in order to devise an efficient pricing model for the Cloud computing scenario.

As services are offered on a subscription basis, they need to be priced based on users' QoS expectations that vary from time to time. It is also important to ensure that whenever service providers are unable to meet all SLAs, their violation needs to be managed to reduce penalties.

---

[6] http://aws.amazon.com/ec2/

Data centers are expensive to operate as they consume huge amount of electricity. For instance, the combined energy consumption of all data centers worldwide is equivalent to the power consumption of Czech Republic. As a result, their carbon footprint on the environment is rapidly increasing. In order to address these issues, energy efficient resource allocation and algorithms need to be developed.

In addition, practical and engineering problems are yet to be solved. Cloud computing infrastructures need to be scalable and reliable. In order to support this, a large number of application service consumers from around the world, Cloud infrastructure providers (i.e., IaaS providers) have been establishing data centers in multiple geographical locations to provide redundancy and ensure reliability in case of site failures. Cloud environments need to provide seamless/automatic mechanisms for scaling their hosted services across multiple, geographically distributed data centers in order to meet QoS expectations of users from different locations. The scaling of applications across multiple-vendor infrastructures requires protocols and mechanisms needed for the creation of InterCloud environments.

From applications' perspective, the development of platform and services that take full advantage of the Cloud Computing model, constitute an interesting software engineering problem.

These are some of the key challenges that need to be addressed for a successful adoption of the Cloud computing paradigm into the mainstream IT industry. R&D initiatives in both academia and industry are playing an important role in addressing these challenges. In particular, the outcome of such research in terms of models, software frameworks, and applications constitute the first tools that can be used to experience Cloud computing. The Cloudbus Toolkit is a step towards this goal.

**2.3 Cloud Computing Reference Model**

Fig. 4 provides a broad overview of the scenario envisioned by Cloud computing. This scenario identifies a reference model into which all the key components are organized and classified. As previously introduced, the novelty of this approach intercepts the entire computing stack: from the system level, where IT infrastructure is delivered on demand, to the user level, where applications transparently hosted in the Cloud are accessible from anywhere. This is the revolutionary aspect of Cloud computing that makes service providers, enterprises, and users completely rethink their experience with IT.

The lowest level of the stack is characterized by the physical resources, which constitute the foundations of the Cloud. These resources can be of different nature: clusters, data centers, and desktop computers. On top of these, the IT infrastructure is deployed and managed. Commercial Cloud deployments are more likely to be constituted by data centers hosting hundreds or thousands of machines, while private Clouds can provide a more heterogeneous environment, in which even the idle CPU cycles of desktop computers are used to leverage the compute workload. This level provides the "horse power" of the Cloud.

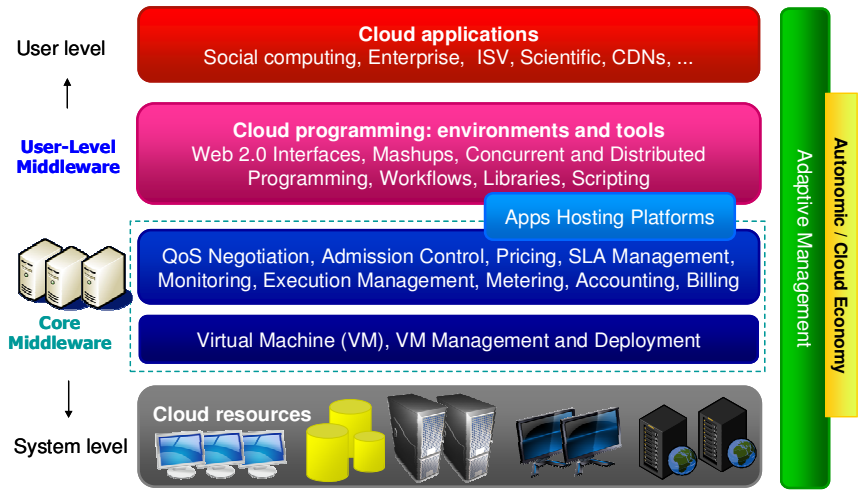

**Fig. 4.** Cloud computing reference model.

The physical infrastructure is managed by the core middleware whose objectives are to provide an appropriate runtime environment for applications and to utilize the physical resources at best. Virtualization technologies provide features such as application isolation, quality of service, and sandboxing. Among the different solutions for virtualization, hardware level virtualization and programming language level virtualization are the most popular. Hardware level virtualization guarantees complete isolation of applications and a fine partitioning of the physical resources, such as memory and CPU, by means of virtual machines. Programming level virtualization provides sandboxing and managed executions for applications developed with a specific technology or programming language (i.e. Java, .NET, and Python). Virtualization technologies help in creating an environment in which professional and commercial services are integrated. These include: negotiation of the quality of service, admission control, execution management and monitoring, accounting, and billing.

Physical infrastructure and core middleware represent the platform where applications are deployed. This platform is made available through a user level middleware, which provides environments and tools simplifying the development and the deployment of applications in the Cloud. They are: web 2.0 interfaces, command line tools, libraries, and programming languages. The user-level middleware constitutes the access point of applications to the Cloud.

At the top level, different types of applications take advantage of the offerings provided by the Cloud computing reference model. Independent software vendors (ISV) can rely on the Cloud to manage new applications and services. Enterprises can leverage the Cloud for providing services to their customers. Other opportunities can be found in the education sector, social computing, scientific computing, and Content Delivery Networks (CDNs).

It is quite uncommon for a single value offering to encompass all the services described in the reference model. More likely, different vendors specialize their business towards providing a specific subclass of services that address the needs of a market sector. It is possible to characterize the different solutions into three main classes: Software as a Service (SaaS), Platform as a Service (PaaS), and Infrastructure / Hardware as a Service (IaaS/HaaS). Table 1 summarizes the nature of these categories and lists some major players in the field.

**Table 1.** Cloud computing services classification.

| Category | Characteristics | Product Type | Vendors & Products |
|---|---|---|---|
| SaaS | Customers are provided with applications that are accessible anytime and from anywhere. | Web applications and services (Web 2.0) | SalesForce.com (CRM) Clarizen.com (Project Management) Google Documents, Google Mail (Automation) |
| PaaS | Customers are provided with a platform for developing applications hosted in the Cloud. | Programming APIs and frameworks; Deployment system. | Google AppEngine Microsoft Azure Manjrasoft Aneka |
| IaaS/HaaS | Customers are provided with virtualized hardware and storage on top of which they can build their infrastructure. | Virtual machines management infrastructure, Storage management | Amazon EC2 and S3; GoGrid; Nirvanix |

Infrastructure as a Service (IaaS) or Hardware as a Service (HaaS) solutions deliver IT infrastructure based on virtual or physical resources as a commodity to customers. These resources meet the end user requirements in terms of memory, CPU type and power, storage, and, in most of the cases, operating system as well. Users are billed on a pay-per-use basis. They have to set up their applications on top of these resources that are hosted and managed in data centers owned by the vendor. Amazon is one of the major players in providing IaaS solutions. Amazon Elastic Compute Cloud (EC2) provides a large computing infrastructure and a service based on hardware virtualization. By using Amazon Web Services, users can create Amazon Machine Images (AMIs) and save them as templates from which multiple instances can be run. It is possible to run either Windows or Linux virtual machines, for which the user is charged per hour for each of the instances running. Amazon also provides storage services with the Amazon Simple Storage Service (S3)[7], users can use Amazon S3 to host large amount of data accessible from anywhere.

Platform as a Service solutions provide an application or development platform in which users can create their own application that will run on the Cloud. More precisely, they provide an application framework and a set of API that can be used by developers to program or compose applications for the Cloud. PaaS solutions often integrate an IT infrastructure on top of which applications will be executed. This is the case of Google AppEngine and Microsoft Azure, while other solutions, such as Manjrasoft Aneka, are purely PaaS implementations.

---

[7] http://aws.amazon.com/s3/

Google AppEngine[8] is a platform for developing scalable web applications that run on top of data centers maintained by Google. It defines an application model and provides a set of APIs that allow developers to take advantage of additional services such as Mail, Datastore, Memcache, and others. AppEngine manages the execution of applications and automatically scales them up/down as required. Google provides a free but limited service, while utilizes daily and per minute quotas to meter and price applications requiring a professional service. Azure[9] is a cloud service operating system that serves as the development, run-time, and control environment for the Azure Services Platform. By using the Microsoft Azure SDK, developers can create services that leverage the .NET Framework. These services have to be uploaded through the Microsoft Azure portal in order to be executed on top of Windows Azure. Additional services, such as workflow execution and management, web services orchestration, and access to SQL data stores, are provided to build enterprise applications. Aneka [20], commercialized by Manjrasoft, is a pure PaaS implementation and provides end users and developers with a platform for developing distributed applications for the Cloud by using .NET technology. The core value of Aneka is a service oriented runtime environment that is deployed on both physical and virtual infrastructures and allows the execution of applications developed by means of various programming models. Aneka provides a Software Development Kit (SDK) helping developers to create applications and a set of tools for setting up and deploying clouds on Windows and Linux based systems. Aneka does not provide an IT hardware infrastructure to build computing Clouds, but system administrators can easily set up Aneka Clouds by deploying Aneka containers on clusters, data centers, desktop PCs, or even bundled within Amazon Machine Images.

Software as a Service solutions are at the top end of the Cloud computing stack and they provide end users with an integrated service comprising hardware, development platforms, and applications. Users are not allowed to customize the service but get access to a specific application hosted in the Cloud. Examples of SaaS implementations are the services provided by Google for office automation, such as Google Mail, Google Documents, and Google Calendar, which are delivered for free to the Internet users and charged for professional quality services. Examples of commercial solutions are SalesForce.com and Clarizen.com, which provide online CRM (Customer Relationship Management) and project management services, respectively.

## 3 Cloudbus Vision and Architecture

Fig.5 provides a glimpse in the future of Cloud computing. A Cloud marketplace, composed of different types of Clouds such as computing, storage, and content delivery Clouds, will be available to end-users and enterprises.

---

[8] http://code.google.com/appengine/
[9] http://www.microsoft.com/azure/

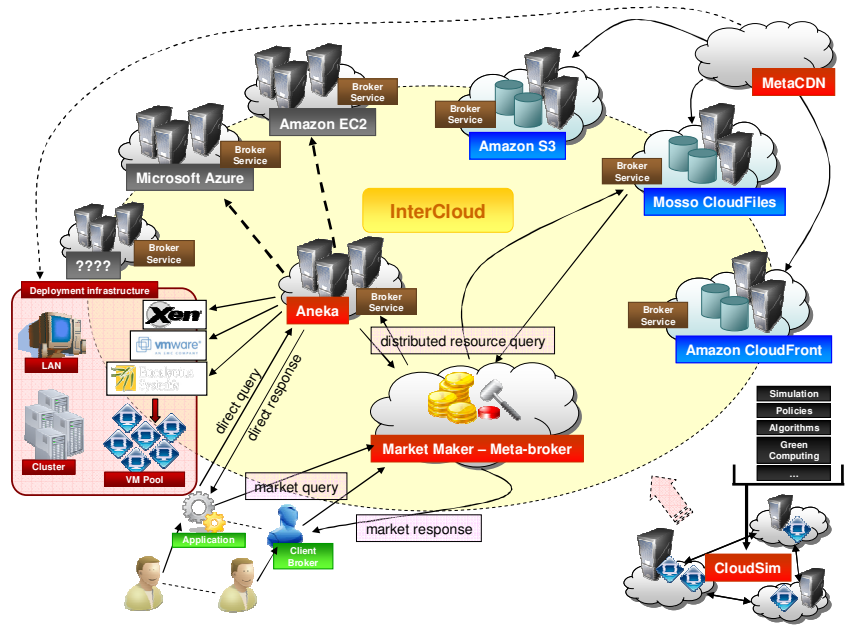

**Fig. 5.** Cloud computing marketplace.

Users can interact with the Cloud market either transparently, by using applications that leverage the Cloud, or explicitly, by making resource requests according to application needs. At present, it is the responsibility of the users to directly interact with the Cloud provider. In the context of a real Cloud marketplace, users will indirectly interact with Cloud providers but they will rely on a *market maker* or *meta-broker* component, which is in charge of providing the best service according to the budget and the constraints of users. A Cloud broker client, directly embedded within applications, or available as a separate tool, will interact with the market maker by specifying the desired Quality of Service parameters through a Service Level Agreement. As a result of the query, the meta-broker will select the best option available among all the Cloud providers belonging to the Cloud marketplace. Such interaction will take place through native interfaces exposed by the provider or via standardized brokering services.

In order to increase their chances of providing a better service to customers, different Cloud providers could establish peering arrangements among themselves in order to offload to (or serve from) other providers' service requests. Such peering arrangements will define a Cloud federation and foster the introduction of standard interface and policies for the interconnection of heterogeneous Clouds. The integration of different technologies and solutions into a single value offering will be the key to the success of the Cloud marketplace. PaaS solutions, such as Aneka [20], could rely on different providers for leveraging the workload and balance the use of private resources by provisioning virtual resources from public Clouds. This approach not only applies for compute intensive services, but also for storage and content delivery. MetaCDN [8], which is a Content Delivery Cloud, aims to provide a unified access to different storage Clouds in order to deliver a better service to end-users and maximize its utility.

The scenario projected by using the Cloud marketplace has its own challenges. Some of them have been already discussed in Section 2.2. In order to make this vision a reality, considerable amount of research has to be carried out through vigorous experiments. Simulation environments will definitely help researcher in conducting repeatable and controllable experiments, while devising new policies and algorithms for resource provisioning or new strategies for an effective and energy efficient use of physical resources. Simulation toolkits, should be able to model any possible scenario and any layer of the Cloud computing reference model: from the fundamental components of the infrastructure, such as physical nodes, data centers, and virtual machines, to the high level services offered to end users. This will help researchers to finely reproduce their problem frame they want to solve and to obtain reliable results.

The Cloudbus Toolkit is a collection of technologies and components that comprehensively try to address the challenges involved in making this vision a concrete reality. Fig. 6 provides a layered view of the entire toolkit and puts it into the context of a real Cloud marketplace. At the top of the stack, real life applications belonging to different scenarios (finance, science, education, engineering, multimedia, and others) leverage the Cloud horse power. Resources available in the Cloud are acquired by means of third party brokering services that mediate the access to the real infrastructure. The Cloudbus toolkit mostly operates at this level by providing a service brokering infrastructure and a core middleware for deploying applications in the Cloud. For what concerns the brokering service, the Market maker is the component that allows users to take full advantage of the Cloud marketplace. The Market maker relies on different middleware implementations to fulfill the requests of users: these can be Cloudbus technologies or third parties implementations. Fig. 6 provides a breakdown of the components that constitute the Cloudbus middleware. Technologies such as Aneka or Workflow Engine provide services for executing applications in the Cloud. These can be public Clouds, private intranets, or data centers that can all be uniformly managed within an InterCloud realm.

In the following sections, we will present more details about the Cloudbus toolkit initiative and describe how they can integrate with each other and existing technologies in order to realize the vision of a global Cloud computing marketplace.

## 4 Cloudbus / CLOUDS Lab Technologies

The CLOUDS lab has been designing and developing Cloud middleware to support science, engineering, business, creative media, and consumer applications on Clouds. A summary of various Cloudbus technologies is listed in Table 2. We briefly describe each of these technologies in the following sub-sections.

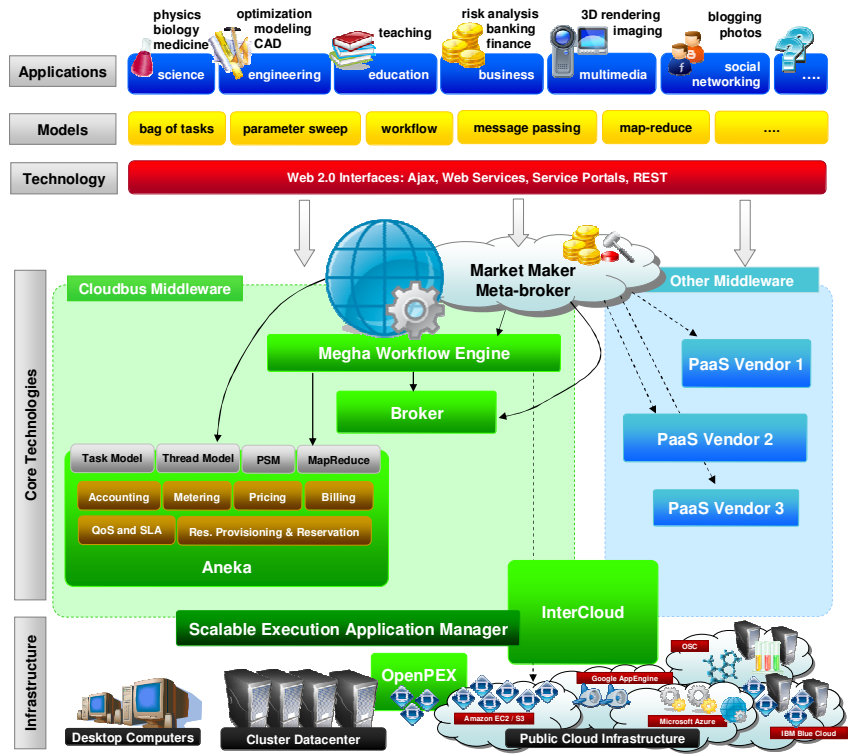

**Fig. 6.** The Cloudbus Toolkit. The picture represents a layered view of the collection of technologies and components for market oriented Cloud computing available within the Cloudbus Toolkit.

### 4.1 Aneka

Aneka [20] is a "Platform as a Service" solution for Cloud computing and provides a software platform for developing and deploying applications in the Cloud. The core features of Aneka are: a) a configurable software container constituting the building blocks of the Cloud; b) an open ended set of programming models available to developers to express distributed applications; c) a collection of tools for rapidly prototyping and porting applications to the Cloud; d) a set of advanced services that put the horse power of Aneka in a market oriented perspective.

One of the elements that make Aneka unique is its flexible design and high level of customization allowing it to target different application scenarios: education, engineering, scientific computing, and financial applications. The Aneka container, which is the core of the component of any Aneka based Cloud, can be deployed into any computing resource connected to the Internet whether it be physical or virtual. This makes the integration with public and private Clouds transparent; and specific services for dynamic provisioning of resources are built into the framework in order to exploit the horse power of the Cloud.

A collection of standardized interfaces, such as Web Services, make Aneka completely integrate with client applications and third party brokering services that can negotiate the desired Quality of Service and submit applications to Aneka Clouds.

**Table 2.** Components of Cloudbus Toolkit.

| Technology | Description |
|---|---|
| **Aneka** | A software platform for developing and deploying Cloud computing applications. |
| **Broker** | A middleware for scheduling distributed applications across Windows and Unix-variant distributed resources. |
| **Workflow Management System** | A middleware that handles dependent tasks, implements scheduling algorithms and manages the execution of applications on distributed resources. |
| **Market Maker/ Meta-Broker** | A matchmaker that matches user's requirements with service providers' capabilities at a common marketplace. |
| **InterGrid** | A model that links islands of Grids through peering arrangements to enable inter-Grid resource sharing. |
| **MetaCDN** | A system that intelligently places users' content onto "Storage Cloud" resources based on their QoS and budget preferences. |
| **Energy Efficient Computing** | A research on developing techniques and technologies for addressing scalability and energy efficiency. |
| **CloudSim** | A simulation toolkit that helps users model: compute, storage, network and other related components of Cloud data centers. |

### 4.2 Broker

The Grid Service Broker [9] mediates access to distributed physical and virtual resources by (a) discovering suitable data sources for a given analysis scenario, (b) selecting suitable computational resources, (c) optimally mapping analysis jobs to compute resources, (d) deploying and monitoring job execution on selected resources, (e) accessing data from local or remote data source during job execution and (f) collating and presenting results. It provides a platform on which enhanced resource brokering strategies can be developed and deployed.

The broker supports various application models such as parameter sweep, workflow, parallel and bag of tasks. It has plug-in support for integration with other middleware technologies such a Globus [21], Aneka [20], Unicore [22] and ssh plug-in for accessing Condor [23], Unix based platforms via fork, PBS [24] and SGE [25]. The broker can provision compute and storage services in Cloud resources via SSH. It also provides QoS parameters in its service description for applications requiring a mix of public and private Cloud resources. For e.g. part of an application workload can be offloaded to Amazon EC2 and rest to local resources dynamically.

### 4.3 Workflow Engine

The Workflow Management System (WMS) [10] aids users by enabling their applications to be represented as a workflow and then execute on the Cloud from a higher level of abstraction. The WMS provides an easy-to-use workflow editor for application composition, an XML-based workflow language for structured representation, and a user-friendly portal with discovery, monitoring, and scheduling components. It can be used with either Aneka [20] and/or Broker [9] to manage applications running on distributed resources. These tools put together enables users to select distributed resources on Clouds and/or Grids, upload/download huge amount of data to/from selected resources, execute applications on distributed resources using various scheduling algorithms and monitor applications' progress in real-time.

The WMS has been used for several real-world applications such as: fMRI brain imaging analysis [11,10], evolutionary multi-objective optimizations using distributed resources [11] and intrusion detection systems with various models [12].

### 4.4 Market Maker/Meta-broker

Market Maker/Meta-broker [13,14] is a part of Cloud infrastructure that works on behalf of both Cloud users and Cloud service providers. It mediates access to distributed resources by discovering suitable Cloud providers for a given user application and attempts to optimally map users' jobs and requirements to published services. It is a part of a global marketplace where service providers and consumers join to find suitable match for each other. It provides various services to its customers such as resource discovery, meta-scheduler, reservation service, queuing service, accounting and pricing services.

### 4.5 From InterGrid to InterCloud

In the coming years, users will be able to see a plethora of Cloud several providers around the world desperate to provide resources such as computers, data, and instruments to scale science, engineering, and business applications. In the long run, these Clouds may require sharing its load with other Cloud service providers as users may select various Cloud services to work on their applications, collectively. Therefore, dispersed Cloud initiatives may lead to the creation of disparate Clouds with little or no interaction between them. The InterCloud model will: (a) promote interlinking of islands of Clouds through peering arrangements to enable inter-Cloud resource sharing; (b) provide a scalable structure for Clouds that allow them to interconnect with one another and grow in a sustainable way; (c) create a global Cyberinfrastructure to support e-Science and e-Business applications.

At present, the InterGrid project [15] is a first step towards realizing the InterCloud vision. It has been implemented using the existing Grid infrastructure. The system uses virtual machines as building blocks to construct execution environments that span multiple computing sites. The computing sites could be a combination of physical machines hosted on Grid sites or virtual machines running on cloud infrastructures, such as Amazon EC2.

### 4.6 MetaCDN

MetaCDN [8] is a system that exploits "Storage Cloud" resources offered by multiple IaaS vendors, thus creating an integrated overlay network that provides a low cost, high performance CDN for content creators. It removes the complexity of dealing with multiple storage providers, by intelligently matching and placing users' content onto one or many storage providers based on their quality of service, coverage and budget preferences. By using a single unified namespace, it helps users to harness the performance and coverage of numerous "Storage Clouds".

### 4.7 Energy Efficient Computing

In order to support elastic applications, Cloud infrastructure providers are establishing Data Centers in multiple geographic locations. These Data Centers are expensive to operate since they consume significant amount of electric power. For instance, the energy consumption of Google Data Center is equivalent to the power consumption of cities such as San Francisco. This is not only increasing the power bills, but also contributing to global warming due to its high carbon footprint. Indeed, the ICT sector is currently responsible for about 2 percent of global greenhouse gas emissions.

In our current research, we are investigating and developing novel techniques and technologies for addressing challenges of: application scalability and energy efficiency with the aim of making a significant impact on industry producing service-oriented Green ICT technologies. As part of this, we explored power-aware scheduling [16], which is one of the ways to reduce energy consumption when using large data-centers. Our scheduling algorithms select appropriate supply voltages of processing elements to minimize energy consumption. As energy consumption is optimized, operational cost decreases and the reliability of the system increases.

### 4.8 CloudSim

The CloudSim toolkit [17] enables users to model and simulate extensible Clouds as well as execute applications on top of Clouds. As a completely customizable tool, it allows extension and definition of policies in all the components of the software stack. This makes it suitable as a research tool as it can relieve users from handling the complexities arising from provisioning, deploying, configuring real resources in physical environments.

CloudSim offers the following novel features: (i) support for modeling and simulation of large scale Cloud computing infrastructure, including data centers on a single physical computing node; and (ii) a self-contained platform for modeling data centers, service brokers, scheduling, and allocations policies. For enabling the simulation of data centers, CloudSim provides: (i) virtualization engine, which aids in creation and management of multiple, independent, and co-hosted virtualized services on a data center node; and (ii) flexibility to switch between space-shared and time-shared allocation of processing cores to virtualized services. These features of CloudSim would speed up the development of new resource allocation policies and scheduling algorithms for Cloud computing.

CloudSim evolved from GridSim [18], a Grid simulation toolkit for resource modeling and application scheduling for parallel and distributed computing. GridSim provides a comprehensive facility for creating different classes of heterogeneous resources that can be aggregated using resource brokers for solving compute and data intensive applications. It provides a framework for incorporating failures, advance reservations, allocation policies, data models, network model extensions, background traffic and load, and so forth, which are also present in the CloudSim toolkit.

## 5 Related Technologies, Integration, and Deployment

The Cloudbus toolkit provides a set of technologies completely integrated with each other. More importantly, they also support the integration with third party technologies and solutions. Integration is a fundamental element in the Cloud computing model, where enterprises and end-users offload their computation to third party infrastructures and access their data anytime from anywhere in a ubiquitous manner.

Many vendors provide different solutions for deploying public, private, and hybrid Clouds. At the lowest level of the Cloud computing reference model, virtual server containers provide a management layer for the commodity hardware infrastructure: VMWare[10], Xen [7], and KVM[11] (Kernel-based Virtual Machine) are some of the most popular hypervisors available today. On top of these, "Infrastructure as a Service" solutions such as Amazon EC2, Eucalyptus [28], and OpenNebula [29] provide a high level service to end-users. Advanced resource managers such as OpenPEX [27] complete the picture by providing an advance reservation based approach for provisioning virtual resources on such platforms. Technologies such as Aneka and the Workflow Engine can readily be integrated with these solutions in order to utilize their capabilities and scale on demand. This applies not only for compute type workloads, but also for storage Clouds and CDNs, as demonstrated by the MetaCDN project. At a higher level, the Market maker and the Grid Service Broker are able to pro-vision compute resources with or without a SLA by relying on different middleware implementations and provide the best suitable service to end-users.

---

[10] http://www.vmware.com
[11] http://www.linux-kvm.org

The Cloudbus toolkit is a work in progress, but several Cloudbus technologies have been already put into action in real scenarios. A private Aneka Cloud has been deployed at GoFront[12] in order to increase the overall productivity of product design and the return of investment of existing resources. The Workflow Engine has been used to execute complex scientific applications such as functional Magnetic Resonance Imaging (fMRI) workflows on top of hybrid Clouds composed of EC2 virtual resources and several clusters in the world [10, 11]. Various external organizations, such as HP Labs are using CloudSim for industrial Cloud computing research.

Furthermore, Aneka has been extended to support dynamic pooling of resources from public Clouds. This capability of Aneka enables creation of hybrid Clouds by leasing additional resources from external/public Clouds such as Amazon EC2 whenever the demand on private Cloud exceeds its available capacity. In addition, Aneka supports federation of other private Clouds within an enterprise, which are managed through Aneka or other vendor technologies such as XenServer and VMWare.

Moreover, some of our Cloudbus technologies have been utilized by commercial enterprises and they are demonstrated at public international events such as the 4th IEEE International Conference on e-Science held in Indianapolis, USA; and the 2nd IEEE International Scalable Computing Challenge hosted at the 9th International Conference on Cluster Computing and Grid (CCGrid 2009) held in Shanghai, China. These demonstrations included fMRI brain imaging application workflows [11, 19], and gene expression data classification [26] on Clouds and distributed resources.

## 6   Future Trends

In the next two decades, service-oriented distributed computing will emerge as a dominant factor in shaping the industry, changing the way business is conducted and how services are delivered and managed. This paradigm is expected to have a major impact on service economy, which contributes significantly towards GDP of many countries, including Australia. The service sector includes health services (e-health), financial services and government services. With the increased demand for delivering services to a larger number of users, providers are looking for novel ways of hosting their application services in Clouds at lower cost while meeting the users' quality of service expectations. With increased dependencies on ICT technologies in their realization, major advances are required in Cloud Computing to support elastic applications offering services to millions of users, simultaneously.

Software licensing will be a major hurdle for vendors of Cloud services when proprietary software technologies (e.g. Microsoft Windows OS) have to be made available to millions of users via public virtual appliances (e.g. customized images of OS and applications). Overwhelming use of such customized software would lead to seamless integration of enterprise Clouds with public Clouds for service scalability and greater outreach to customers. More and more enterprises would be interested in moving to Clouds for cooperative sharing. In such scenarios, security and privacy of corporate data could be of paramount concern to these huge conglomerates. One of the solutions would be to establish a globally accredited Cloud service regulatory body that would act under a common statute for certifying Cloud service providers, standardizing data formats, enforcing service level agreements, handling trust certificates and so forth.

---

[12] http://www.gofront.com

On one hand, there are technological challenges; on the other, there are issues with balancing usage cost and services delivered. Cloud service providers are already tussling by advertising attractive pricing policies for luring users of all kinds to use their services (e.g. Amazon, SalesForce, Google, etc.). As the market condition is determined through cutthroat competition between many vendors, dynamic negotiations and SLA management will play a major role in determining the amount of revenue to be generated for service providers. Similarly, users will be able to choose better services that fit their requirements and budget. They will be evaluating services based on their level of QoS satisfaction, so that they get the right value for the price paid.

As the price for commodity hardware and network equipments for a data center is already getting cheaper, significant part of the total cost of operating Cloud services in industrial scale is determined by the amount of energy consumed by the data center. To conserve energy and save cooling costs, data centers could adopt energy efficient resource allocation policies. Moreover, they could use renewable sources of energy to power up their centers and leave the least carbon footprint, in the long run.

A daunting task for any vendor is to keep its Cloud services alive and running for as long as it takes. As users gradually become dependent on Cloud services, a sudden disruption of any of the services will send a ripple effect around the world that could: destabilizing markets (e.g. financial institutions such as banks depending on Clouds), paralyzing IT services (e.g. gmail services) and so forth. For preventing these effects arising from vendor "lock-in", interoperability issues between Cloud service providers should be adequately addressed.

Nevertheless, Cloud Computing is the technology for realizing a long awaited dream of using distributed compute, storage resources and application software services as commodities (computing utilities).

As the technology is gradually changing from Cluster and Grid computing to Cloud computing, the Cloudbus toolkit is also evolving towards being more robust and scalable to support the hype. We are continuously consolidating our efforts to enhance the toolkit such that it is able to support more and more users.

**Acknowledgments.** All members of our CLOUDS Lab have been actively contributing towards various developments reported in this paper. In particular, we would like to thank Srikumar Venugopal, Xingchen Chu, Rajiv Ranjan, Chao Jin, Michael Mattess, William Voorsluys, Dileban Karunamoorthy, Saurabh Garg, Marcos Dias de Assunção, Alexandre di Costanzo, Mohsen Amini, James Broberg, Mukaddim Pathan, Chee Shin Yeo, Anton Beloglazov, Rodrigo Neves Calheiros, and Marco Netto.